\documentclass[aps,prl,preprint,groupedaddress]{revtex4}
\begin{document}
\title{\bf  Production and detection of heavy matter anti-matter from Higgs decays }
\author{Y. N. Srivastava}
\affiliation{INFN \& Dipartimento di Fisica, Universita' di Perugia, Perugia, Italy}
\author{A. Widom}
\affiliation{Physics Department, Northeastern University, Boston, MA 02115 USA }
\author{J. Swain}
\affiliation{Physics Department, Northeastern University, Boston, MA 02115 USA }
\date{\today}

\begin{abstract}
The one-loop Higgs coupling to two gluons has been invoked in the past to estimate 
that the fraction of the nucleon mass which is due to the Higgs is rather small but 
calculable (approximately 8 percent). To test the 
veracity of this hypothesis, we  employ the same mechanism to compute the Higgs coupling 
to an arbitrary stable nucleus $A$ and its anti-nucleus $\bar{A}$. We find that the physical 
decay rate of a Higgs into a spin zero $A\bar{A}$ pair near the threshold corresponding 
to the Higgs mass is quite substantial, once we include the final state Coulomb corrections
as well as possible form factor effects. If true, observation of even a few such decay events 
would be truly spectacular (with no competing background) since we are unaware of any other 
interaction which might lead to the production of a very heavy nucleus accompanied by its anti 
nucleus in nucleon-(anti-) nucleon scattering. 
\end{abstract}
\maketitle
\section{Introduction}
In the Standard Model, masses to the elementary fermions (leptons and quarks) and 
the gauge bosons ($W^±$ and $Z^o$) are supposedly provided by the Higgs boson
through the spontaneous symmetry breaking mechanism and all these masses are
proportional to the vacuum expectation value $v$ of the Higgs. Thus, the coupling
needed to compute the Higgs decay into an elementary ($f\bar{f}$) pair (where $f$ denotes
a lepton or a quark) is given by the S-matrix element
\begin{equation}
\label{el1}
< f(p_1,s_1); \bar{f}(p_2,s_2) | S | H(K) > = [(2\pi)^4\delta^4(K - p_1 - p_2)] i ({{m_f}\over{v}})
[\bar{u}(p_1, s_1) v(p_2, s_2)], 
\end{equation}
By contrast, the corresponding coupling of the Higgs to a nucleon which is a composite 
state of light flavor $u$ and $d$ quarks and glue is considered a dynamical question (see
for example the textbook by Okun[1]). The argument used in this computation invokes
the trace anomaly and the fact that the observable mass of a hadron such as a nucleon 
(assumed made of light flavors) receives only a negligible contribution from the masses 
of these quarks since their masses are so tiny. Thus, in the chiral limit, it is sufficient to:
\par\noindent
(i) compute the coupling of the Higgs to two gluons ( which in the one-loop approximation 
is dominated by the heaviest mass quark, the top),
\par\noindent
(ii) relate the trace of the energy momentum tensor of any hadron (made-up of chiral
quarks) to the trace anomaly given by the two gluons, 
\par\noindent
(iii) use the fact that the physical mass of any object is given through the trace of its 
energy momentum  tensor,
\par\noindent
for a computation of the Higgs coupling to the hadron and thus estimate the fraction of  
the physical mass of a hadron which is generated by the Higgs mechanism. To complete
the calculation of the physical decay of the Higgs into a hadronic channel, the above must 
be augumented by Coulomb corrections in the final charged states as well as possible
effects due to the hadronic form factors in continuing the matrix element from $Q^2\ =\ 0$ to
$Q^2\ =\ M_H^2$. 
\par\noindent
It is well to remember that this argument -if applicable at all- should be 
equally applicable to any hadron made of chiral quarks. We reproduce this calculation
below.

Consider first the analog of Eq.(\ref{el1}) above, that is the physical S-matrix element for the 
decay of  a Higgs into a spin $1/2$ hadron and its anti-hadron
\begin{equation}
\label{el2}
S[H(K)\rightarrow A(p_1,s_1)\bar{A}(p_2,s_2)] =  < A(p_1,s_1);\bar{A}(p_2,s_2)| S | H(K)>, 
\end{equation}
Following step(i) outlined above, we insert for the S-operator only the Higgs coupling
to two gluons, i.e.,
\begin{equation}
\label{el3}
S = e^{i\int(d^4x){\cal L}_{Hgg}(x)},
\end{equation}
and take the one-heavy quark loop value for ${\cal L}_{Hgg}(x)$
\begin{equation}
\label{el4}
{\cal L}_{Hgg}(x) = [{{-\alpha_s(M_H)}\over{12\pi}}] ({{\sigma(x)}\over{v}})
G_{\mu\nu}^c(x) G^{\mu\nu}_c(x), 
\end{equation}
where $\alpha_s$ is the QCD running coupling constant, which should be evaluated 
at the Higgs mass when computing the physical Higgs decay into two gluons; $\sigma(x)$ 
is the physical Higgs field and $G_{\mu\nu}^c(x)$ is the gluon field strength tensor 
with $c$ denoting its color. Inserting Eq.(\ref{el1}) into Eq.(\ref{el2}) and employing 
the Born approximation, we find
\begin{equation}
\label{el5}
S[H(K)\rightarrow A(p_1,s_1)\bar{A}(p_2,s_2)]  =[{{-i\alpha_s(M_H)}\over{12\pi v}}]
{\cal M}_{gg}(K),
\end{equation}
where the matrix element coupling the hadron to two gluons ${\cal M}_{gg}(K)$ is
defined via
\begin{equation}
\label{el6}
{\cal M}_{gg}(K) = \int (d^4x) e^{-i K\cdot x} <A(p_1,s_1);\bar{A}(p_2,s_2)|G_{\mu\nu}^c(x) 
G^{\mu\nu}_c(x)|0 >.
\end{equation}  
To implement step(ii), we recall that in the chiral limit, the trace of the energy momentum
tensor pertaining to any hadron made of light flavors only is given entirely by the anomaly
term
\begin{equation}
\label{el7}
 {\hat T}(x)  \approx\  9[{{-\alpha_s(M_H)}\over{8\pi}}] G_{\mu\nu}^c(x) G^{\mu\nu}_c(x),
\end{equation}
where the factor of 9 is for three light flavors. Using Eqs.(\ref{el6}) and (\ref{el7}), 
we obtain
\begin{equation}
\label{el8}
S[H(K)\rightarrow A(p_1,s_1)\bar{A}(p_2,s_2)]  = 
(2i/27 v) \int (d^4x) e^{-i K\cdot x} < A(p_1,s_1);\bar{A}(p_2,s_2)| {\hat T}(x) | 0 >.
\end{equation}
and
\begin{equation}
\label{el9}
S[H(K)\rightarrow A(p_1,s_1)\bar{A}(p_2,s_2)] \approx\ 
 (2i/27 v) {\cal M}_{T}(K), 
\end{equation}
where the matrix element coupling the hadron to the trace of the energy momentum tensor 
${\cal M}_{T}(K)$ is given by
\begin{equation}
\label{el10}
{\cal M}_{T}(K) = \int (d^4x) e^{-i (K- p_1 - p_2)\cdot x} < A(p_1,s_1);\bar{A}(p_2,s_2)| {\hat T}(0) | 0 >.
\end{equation}
Step (iii) relates the above matrix element to the physical mass $M_A$ of the spin $1/2$
hadron $A$
\begin{equation}
\label{el11}
< A(p_1,s_1);\bar{A}(p_2,s_2)| {\hat T}(0) | 0 > = M_A [\bar{u}(p_1, s_1) v(p_2, s_2)],
\end{equation}
and we obtain the sought for result
\begin{equation}
\label{el12}
S[H(K)\rightarrow A(p_1,s_1)\bar{A}(p_2,s_2)]  = 
[(2\pi)^4\delta^4(K - p_1 - p_2)] i ({{2 M_A}\over{27 v}})[\bar{u}(p_1, s_1) v(p_2, s_2)],
\end{equation}
Thus, we find that the (dimensionless) Higgs decay coupling to any light flavored spin $1/2$ 
hadron $A_F$ and its anti-hadron $\bar{A}_F$ is given by $(2/27)(M_F/v)$, where 
$M_F \approx\ A_F M_N$ denotes the mass of a spin-half nucleaus made of $A_F$ nucleons each 
of mass $M_N$. This coupling is not at all negligible for large $A_F$. The Higgs decay width 
into this channel is given by
\begin{equation}
\label{el13}
\Gamma_o(H \rightarrow A_F \bar{A}_F) = [{{1}\over{8 \pi}}][{{2 M_F}\over{27 v}}]^2
[1 - (2M_F/M_H)^2 ]^{3/2} M_H.\ \ (2M_F\ <\ M_H),
\end{equation}
prior to final state Coulomb corrections and possible form factor effects to which we shall 
return momentarily. A similar calculation for the Higgs decay into a pair of spin zero nucleus 
$A_B$ and its anti-nucleus $\bar{A}_B$ gives
\begin{equation}
\label{el14}
\Gamma(H \rightarrow A_B \bar{A}_B) = [{{1}\over{16\pi}}][{{2  M_B^2}\over{27 v M_H}}]^2
[1 - (2 M_B/M_H)^2  ]^{1/2} M_H |f_B(M_H^2; M_B^2)|^2 P_{EM}(Z_B; v), 
\end{equation}
where $2M_B\ <\ M_H$ and $f_B(M_H^2; M_B^2)$ denotes the complex $S$-wave form factor for the 
spin-zero hadron of mass $M_B$ coupling to Higgs evaluated at $M_H^2$, when normalized to be 
$1$ at $Q^2\ =\ 0$, that is when, $f_B(0; M_B^2)\ =\ 1$. On the other hand, $P_{EM}(Z_B; v)$ 
denotes the final state Coulomb corrections between a nucleus of charge ($Z_B e$) and an 
anti-nucleus of charge ($-Z_B e$) with relative velocity $v$ and for which we have the 
well-known expression
\begin{equation}
\label{el15}
P_{EM}(Z_B; v) = {{\gamma}\over{1 - e^{- \gamma}}};\ \ \gamma\ = {{2c\pi Z_B^2 \alpha}\over{v}}.  
\end{equation}
For large $Z_B$ and near threshold $v/c\ <<\ 1$, we may approximate the above
\begin{equation}
\label{el16}
P_{EM}(Z_B >>\ 1; v<<\ c) = {{2\pi Z_B^2 \alpha\over{\sqrt{1 - (2M_B/M_H)^2}}}}.  
\end{equation}   
Hence, near threshold ($M_H\ \approx\ 2M_B$), for a spin-zero pair, the threshold factors 
cancel out between Eq.(\ref{el14}) and Eq.(\ref{el16}) so that the Higgs decay rate remains 
finite. Not so, for spin half, which has a $P$-wave threshold dependence (vedi, Eq.(\ref{el13})).

In modulus, the form-factor is expected to remain close to $1$ near threshold. For a heavy 
nucleus, if the form factor modulus is assumed to be strongly suppressed for $Q^2\ >>\ M_H^2$, 
then dispersion relation considerations impel one to conclude that in order for the form factor 
to attain its value unity at $Q^2\ =\ 0$, the modulus of the function cannot be small near 
threshold. We have corraborative evidence from the experimental measurements of the $S$-wave, 
time-like, EM form factors near their respective baryon-antibaryon thresholds for 
$\gamma^*\ \rightarrow\ p\bar{p}$ and  $\gamma^*\ \rightarrow\ \Lambda_c\bar{\Lambda}_c$. 
Within experimental errors, $|G^p(4M_p^2)|\ =\ 1$ and the same is true replacing the proton
with a $\Lambda_c$\cite{Babar}\cite{Belle}\cite{Zichichi}\cite{Baldini} For the 
phenomenological analysis to follow, we 
shall take for the modulus of the form factor, its nominal value unity.

Hence, both Coulomb and form factor considerations lead us to conclude that the dominant decay 
mode of the Higgs would be into that spin-zero nucleus-antinucleus pair whose individual mass 
$M_B$ would be the closest to $M_H/2$ and with $Z_B$ as large as possible. For illustrative 
purposes only, let us take a spin zero nucleus with an equal number of protons and nucleons, so 
that $Z_B\ =\ A_B/2$ and consider a nucleus with mass $M_B \approx\ M_H/2$. The Higgs decay width 
near threshold under these assumptions is given by
\begin{equation}
\label{el17}
\Gamma(H \rightarrow A_B \bar{A}_B)_{threshold} \approx [{{\alpha}\over{8}}][{{M_H^2}\over{216 m_N v}}]^2 M_H,
\end{equation}  
which should be compared to the Higgs decay channel $b\bar{b}$, generally considered dominant 
(for $M_H\ <\ 2M_W$)
\begin{equation}
\label{el18}
\Gamma(H \rightarrow b \bar{b}) = [{{3}\over{8\pi}}][{{m_b}\over{v}}]^2 M_H, 
\end{equation}   
where $m_b$ is the $b$-quark mass. Defining the ratio
\begin{equation}
\label{el19}
R(M_H) = \big{[}{{\Gamma(H \rightarrow A_B \bar{A}_B)|_{threshold}}\over{\Gamma(H \rightarrow b \bar{b})}}\big{]},
\end{equation}
we find
\begin{equation}
\label{el20}
R(M_H)= [{{\pi \alpha}\over{3}}][{{M_H^2}\over{216 m_N m_b}}]^2,
\end{equation}
where $m_N$ denotes the mass of the nucleon. Using $m_b\ =\ 4.2\ GeV$,\cite{PDG} we find that
the ratio $R$ varies between $3.8$ for $M_H\ =\ 126\ GeV$ to $9.4$ for $M_H\ =\ 160\ GeV$. Clearly, 
this channel dominates the Higgs decay for the entire range of the allowed Higgs mass.

\end{document}